\documentclass[12pt]{article}
\usepackage{amsmath,amssymb,amsfonts}
\usepackage{type1cm}
\usepackage[dvipdfmx]{graphicx}
\usepackage{booktabs}
\renewcommand{\theequation}{\thesection.\arabic{equation}}
\renewcommand{\thefootnote}{\fnsymbol{footnote}}
\newcommand{\beq}{\begin{equation}}
\newcommand{\eeq}{\end{equation}}
\newcommand{\bea}{\begin{eqnarray}}
\newcommand{\ena}{\end{eqnarray}}
\newcommand{\ba}{\begin{align}}
\newcommand{\ea}{\end{align}}
\newcommand{\vs}[1]{\vspace{#1 mm}}

\newcommand{\uda}{\nearrow \kern-1em \searrow}
\makeatletter
\def\eqnarray{%
 \stepcounter{equation}%
 \let\@currentlabel=\theequation
 \global\@eqnswtrue
 \global\@eqcnt\z@
 \tabskip\@centering
 \let\\=\@eqncr
 $$\halign to \displaywidth\bgroup\@eqnsel\hskip\@centering
 $\displaystyle\tabskip\z@{##}$&\global\@eqcnt\@ne
 \hfil$\displaystyle{{}##{}}$\hfil &\global\@eqcnt\tw@$\displaystyle\tabskip\z@{##}$\hfil
 \tabskip\@centering&\llap{##}\tabskip\z@\cr}
\makeatother
\def\l{\langle}\def\p{\partial}

\begin{document}
\begin{titlepage}
\setcounter{page}{0}
\begin{flushright}
EPHOU-15-019\\
\today\\
\end{flushright}

\vs{6}
\begin{center}
{\Large \textbf{ Mirror Symmetry of Minimal Calabi-Yau Manifolds
}}

\vs{6}
\textbf{Hideyuki Kawada$^{1}$, \ Takahiro Masuda$^{2}$ \ and \ 
Hisao Suzuki$^{1}$}\\
{\small \em ${}^1$ Department of Physics, 
Hokkaido University, Sapporo 060-0810, Japan \\
${}^2$ Hokkaido University of Science, Sapporo 006-8585, Japan}

\end{center}
\vs{6}

\centerline{{\bf{Abstract}}}
We perform the mirror transformations of Calabi-Yau manifolds with one moduli 
whose Hodge numbers $(h^{11}, h^{21})$ 
are minimally small. 
Since the difference of Hodge numbers is the generation 
of matter fields in superstring theories made of compactifications, 
minimal Hodge numbers of the model of phenomenological interest are (1,4). 
Genuine minimal Calabi-Yau manifold which has least 
degrees of freedom for K\"ahler and complex deformation is (1,1) model. 
With help of {\it Mathematica} and  {\it Maple}, we derive Picard-Fuchs equations 
for periods, and determine their monodromy behaviors completely 
such that all monodromy matrices are consistent 
in the mirror prescription of the model (1,4), (1,3) and (1,1). 
We also discuss to find the description for each mirror of (1,3) and (1,1) 
by combining invariant polynomials of variety on which (1,5) model 
is defined. 
The genus 0 instanton numbers coming from mirror transformations 
in above models look reasonable. 
We propose the weighted discriminant for genus 1 
instanton calculus which makes all instanton numbers integral, except 
(1,1) case. 

\end{titlepage}
\newpage

\renewcommand{\thefootnote}{\arabic{footnote}}
\setcounter{footnote}{0}

\tableofcontents
\section{Introduction}
A Calabi-Yau manifold is partially characterized by 
the Hodge numbers ($h^{11}, h^{21}$). These are topological numbers 
which count the number of parameters that deform the K\"ahler
class and the complex structure of the manifold. 
Recently, Calabi-Yau manifolds are paid attention where 
both Hodge numbers ($h^{11},\  h^{21}$) are small \cite{COHS, CD, BCD}. Many 
Calabi-Yau manifolds with various Hodge numbers
are provided by construction with 
hypersurfaces in weighted projective spaces or in toric varieties. 
However manifolds admitting freely-acting discrete 
symmetry seem to be rare \cite{SW}. Models with 
small Hodge numbers have been found to classify all the freely 
acting symmetries for the 
manifolds \cite{CD, Br2, CC, Br, Dav, Dav2}. Besides the way of
constructions, the Hodge numbers for the models with symmetry of order four
have been calculated recently \cite{CCM}. 

Phenomenologically, it is interesting to search 
string theories with three generations compactified on 
Calabi-Yau manifolds with small Hodge numbers. 
Especially model (1,4) with $\chi=-6$ is the minimal theory 
which have been discussed in \cite{BCD, BCDD}. 
Theoretically, it is worth investigating the case with  
minimal Hodge numbers $(1,1)$ found in \cite{Br}, whose enumerative property 
is not clear so far. 
These models are made by taking quotient of freely acting symmetry
groups, so that Hodge numbers become both small \cite{CD, BCD, SW, Br}, 
however their defining equations turn out to be complicated. 
In these cases, it is not obvious to carry out ordinary 
systematic calculation to derive Picard-Fuchs equation, 
and to perform the mirror transformation to calculate the 
instanton corrections. 

In this paper, 
we attend to investigate the mirror transformations 
for one moduli models with small Euler numbers $|\chi|\leq 8$ 
with aid of computer algebra systems {\it Mathematica} and 
{\it Maple}. Using {\it Mathematica} package ``Generationgfunctions'',
we derive the Picard-Fuchs equation in such models. 
Monodromy behaviors are determined by 
numerical integration on {\it Maple} \cite{ES}. 
In order to determine the symplectic basis of periods, as well as, 
topological indices of such models, we evaluate 
bi-linear form on periods numerically \cite{MS, HK}. 
To check the consistency of the results, 
we calculate the genus 0 and 1 instanton numbers to be integral 
values.  Also it is interesting to investigate 
the relation between the models with $(1,h^{21})$ where $h^{21}<6$, 
and their mirrors. 

Picard-Fuchs equations for periods of Calabi-Yau three-folds have been 
studied extensively, and many kinds of equations have been found already 
in physical or mathematical contexts. 
Restricted to one moduli case, the automated search for 4th order 
Picard-Fuchs equations of Calabi-Yau type with maximally unipotent 
monodromy have been carried out\cite{AESZ}, and 
vast results including previously found operators 
have been summarized in ``Calabi-Yau Operators  
Database'' on the web site \cite{AESZ}.   
Picard-Fuchs equations we derive in this paper 
are found in the database.

In section 2, we review mirror model of (1,4) which is known example. 
From the evaluation of periods, we find Picard-Fuchs equation. 
Bi-linear form on periods gives the characteristic numbers, 
such as Euler number, Yukawa coupling $K$, $c_2$, 
to suggest the way to determine the basis 
of monodromy.  Instanton calculations of genus 1 as well as genus 0 
are performed by using mirror map so that all instanton numbers becomes 
integral, where we propose weighted discriminant for 
1-loop level determined from behaviors around singular points.
In section 3, 
we investigate the sequence of manifolds with 
small Hodge numbers such as  mirror models of (1,5), (1,3), (1,1). 
Starting from the invariant polynomials for (1,5) model, 
and choosing suitable combinations 
of them, we propose the definition of 
mirror models of (1,5), (1,3), (1,1) respectively. 
The mirror transformations in these models can be carried out in 
similar ways in section 2. 
The results about monodromy behaviors and 
instanton numbers are all consistent, except that 
instanton numbers for minimal model (1,1) look strange. 

\section{Minimal model for three generations}
\label{(4,1)}

As an examples of the Calabi-Yau manifold with small Hodge 
numbers ($h^{11},h^{21}$), 
we investigate the model with (1,4), and its mirror, which were 
found in \cite{CD, BCD}. 
This Calabi-Yau manifold is constructed from 
$X^{8,44}$, and was found 
in the course of the project to classify all the freely 
acting symmetries for the manifolds of the CICY list. 
Original space 
$X^{8,44}$ has Euler number $-72$, and is invariant under 
freely acting group $G$ whose order is $12$. 
As is explained in \cite{BCD}, quotient variety $X^{8,44}/G$ is smooth and 
has Euler number $\chi=-72/12=-6$. 
The definition of this model consists of following 
three curves on six manifolds 
\begin{align}
p&=1+s_0s_1+s_1s_2+s_2s_0,\ \ q=1+t_0t_1+t_1t_2+t_2t_0, \notag\\
r&=s_0s_1s_2t_0t_1t_2+
c_1(s_0t_0+s_1t_1+s_2t_2)+c_2(s_0t_1+s_1t_2+
s_2t_0)\\
&+c_3(s_0t_2+s_1t_0+s_2t_1)
+c_4(s_1s_1s_2(t_0+t_1+t_2)+(s_0+s_1+s_2)t_0t_1t_2),\notag
\end{align}
where $c_i$ are four kinds of moduli parameters. 

This is a minimal model of string theory with 
three generations compactified on 
Calabi-Yau manifold with $\chi=-6$. 
Phenomenological aspects about of this model were discussed 
in detail in \cite{BCD, BCDD}.

\subsection{Mirror prescription}

Toric description of this model is also given in \cite{BCD}, and 
alternative defining curve consists of four parameter family of 
invariant Laurent polynomials in terms of homogeneous coordinates 
made of polyhedron $\Delta$ as
\begin{gather}
f=1+\sum_{i=1}^4\gamma_iQ_i
\end{gather} 
where 
\begin{align}
Q_1&=t_1+\frac{1}{t_1}+t_2+\frac{1}{t_2}+t_3+\frac{1}{t_3}+t_4+
\frac{1}{t_4}+\frac{t_1}{t_2}+\frac{t_2}{t_1}+\frac{t_3}{t_4}+
\frac{t_4}{t_3}, \notag\\
Q_2&=(t_1+\frac{1}{t_1})(t_3+\frac{1}{t_3})+
(t_4+\frac{1}{t_4})(\frac{t_1}{t_2}+\frac{t_2}{t_1})+
(t_2+\frac{1}{t_2})(\frac{t_3}{t_4}+\frac{t_4}{t_3}),\notag\\
Q_3&=(t_1+\frac{1}{t_1})(t_4+\frac{1}{t_4})+
(t_2+\frac{1}{t_2})(t_3+\frac{1}{t_3})+
(\frac{t_1}{t_2}+\frac{t_2}{t_1})(\frac{t_3}{t_4}+\frac{t_4}{t_3}),\\
Q_4&=(t_2+\frac{1}{t_2})(t_4+\frac{1}{t_4})+
(t_3+\frac{1}{t_3})(\frac{t_1}{t_2}+\frac{t_2}{t_1})+
(t_1+\frac{1}{t_1})(\frac{t_3}{t_4}+\frac{t_4}{t_3}).\notag
\end{align} 

Obtaining the dual $\nabla$ is to delete the vertices of $\Delta$, which 
corresponds to setting parameters $\gamma_i$ equal to zero except one. 
One of defining curve for the mirror of (1,4) model is \cite{BCD}
\begin{gather}
 P=1+\gamma_1(t_1+\frac{1}{t_1}+
t_2+\frac{1}{t_2}+t_3+\frac{1}{t_3}+t_4+\frac{1}{t_4}+
\frac{t_1}{t_2}+\frac{t_2}{t_1}+\frac{t_3}{t_4}+\frac{t_4}{t_3}).
\end{gather} 
In ordinary cases of toric varieties,  
you may find the  Picard-Fuchs equation for 
the period integral $\int\Pi \frac{dt_i}{t_i}\frac{1}{P}$ following 
the Griffiths-Dwork method. 
Differently it seems difficult to do in this case. 
So we first turn to find an exact form of 
the fundamental period $\omega_0$ by picking up simple 
poles of period integral. 
Residues at $t_1=t_2=t_3=t_4=0$ are calculated by 
expanding $1/P$ as
\begin{gather}
\frac{1}{P}=\sum
\begin{pmatrix}n\\i\end{pmatrix}
\left(t_1+\frac{1}{t_1}+t_2+\frac{1}{t_2}+
\frac{t_1}{t_2}+\frac{t_2}{t_1}\right)^i
\left(t_3+\frac{1}{t_3}+t_4+\frac{1}{t_4}+
\frac{t_3}{t_4}+\frac{t_4}{t_3}\right)^{n-i}z^n.
\end{gather} 
Fundamental period is 
\begin{gather}
\omega_0=\sum
\begin{pmatrix}n\\i\end{pmatrix}
\begin{pmatrix}i\\l_1\end{pmatrix}  
\begin{pmatrix}i-l_1\\l_2\end{pmatrix}  
\begin{pmatrix}l_1\\k_1\end{pmatrix}  
\begin{pmatrix}l_2\\k_2\end{pmatrix}  
\begin{pmatrix}i-l_1-l_2\\k_3\end{pmatrix} \nonumber\\
\cdot \begin{pmatrix}n-i\\p_1\end{pmatrix}  
\begin{pmatrix}n-i-p_1\\l_2\end{pmatrix}  
\begin{pmatrix}p_1\\m_1\end{pmatrix}  
\begin{pmatrix}p_2\\m_2\end{pmatrix}  
\begin{pmatrix}n-i-p_1-p_2\\m_3\end{pmatrix} z^n
\end{gather}
where
\begin{gather}
l_1=i-2k_3-2k_2,\ l_2=i-2k_1-2k_2,\notag\\
p_1=n-i-2m_3-2m_2,\ p_2=n-i-2m_1-2m_2.
\end{gather} 
The multiple summations look still hard to 
derive Picard-Fuchs equation. 
Then we have recourse to  the power of
computer. 
By using {\it Mathematica} package ``Generationgfunctions'', we can find  
a differential equation for series expanded function. First we expand 
$w_0$ in {\it Mathematica} up to high 
enough orders, such as $O(z^{70 })$. Next we apply 
the command ``GuessRE'' which derive the recursion equation among the
coefficients of series expansion. 
After deriving the recursion equation, the command ``RE2DE'' 
tells us the differential equation for this period. The result is 
\begin{gather}
a_4(z)\frac{\partial^4 f}{\partial z^4} 
+a_3(z)\frac{\partial^3 f}{\partial z^3} 
+a_2(z)\frac{\partial^2 f}{\partial z^2}
+a_1(z)\frac{\partial f}{\partial z}
+a_0(z)f=0 
\end{gather} 
where
\begin{align}
a_4(z)=&-z^3\,\left(2\,z-3\right)^2\,\left(3\,z-1\right)\,\left(4\,z-1
 \right)\,\left(4\,z+1\right)\,\left(5\,z+1\right)\,\left(6\,z+1
 \right)\,\left(12\,z-1\right),\notag\\
a_3(z)=&-2\,z^2\,\left(2\,z-3\right)\,(276480\,z^7-478656\,z^6-11232\,
 z^5+55844\,z^4+1100\,z^3\notag \\
&-1701\,z^2-52\,z+9),\notag\\
a_2(z)=&-z\,(4976640\,z^8-16982784\,z^7+14544576\,z^6+880992\,z^5-
 1286856\,z^4\notag \\&-29468\,z^3+26098\,z^2+555\,z-63), \\
a_1(z)=&-(6635520\,z^8-23846400\,z^7+22194432\,z^6+610656\,z^5-1445856
 \,z^4\notag \\&
-12968\,z^3+17532\,z^2+156\,z-9),\notag\\
a_0(z)=&-48\,z\,\left(34560\,z^6-130464\,z^5+132120\,z^4+284\,z^3-6182\,z^2
 +9\,z+36\right).\notag
\end{align} 
This Picard-Fuchs equation is found in the database on the web site 
\cite{AESZ}, though corresponding record number is not manifest there. 
For later convenience, we define the ratio of coefficients as 
\begin{gather}
 r_3(z)=\frac{a_3(z)}{a_4(z)},\ r_2(z)=\frac{a_2(z)}{a_4(z)}, \ 
r_1(z)=\frac{a_1(z)}{a_4(z)}. 
\end{gather}
Classical Yukawa coupling  
\begin{gather}
K_c[z]=\frac{6(3 - 2 z)}{(1 - 3 z)(1 - 4 z)(1 + 4 z)(1 + 5 z)(1 + 6 z)(1 - 12
z)}
\end{gather} 
is basically a quantity $\exp(-\frac{1}{2}\int r_3(z)\,dz)$. 
To check the consistency for Calabi-Yau, we see that 
\begin{gather}
\frac{1}{2}r_2(z)r_3(z)-\frac{1}{8}
r_3(z)^3+r_2'(z)-\frac{3}{4}r_3(z)r_3'(z)-\frac{1}{2}
r_3''(z)-r_1(z)=0
\end{gather} 

The local property of the solutions of Picard-Fuchs equation is
summarized the $P$ symbol as follows. 
\begin{align}
\left\{
\begin{array}{ccccccccc}
-1/4&-1/5&-1/6&0&1/12&1/4&1/3&3/2&\infty\\
\hline
0&0&0&0&0&0&0&0&1\\
1&1&1&0&1&1&1&1&2\\
1&1&1&0&1&1&1&3&3\\
2&2&2&0&2&2&2&4&4
\end{array}
\right\}
\end{align} 
From Picard-Fuchs equation, we can have other three independent
solutions besides $\omega_0$
\begin{align}
\omega_1=&\log z\cdot \omega_0+\Omega_1(z),\notag\\
\omega_2=&(\log z)^2\cdot \omega_0+2\log z\cdot \Omega_1(z)
+\Omega_2(z),\\
\omega_3=&(\log z)^3\cdot \omega_0+3(\log z)^2\cdot 
\Omega_1(z)+3\log z\cdot \Omega_2(z)+
\Omega_3(z),\notag
\end{align}
where polynomial part of above solutions 
$\Omega_1, \Omega_2, \Omega_3$ are obtained order by 
order as
\begin{align}
\Omega_1(z)=&z+\frac{31}{2}z^2+\cdots \notag\\
\Omega_2(z)=&\frac{2}{3}z+\frac{31}{6}z^2+\cdots \\
\Omega_3(z)=&-4z-\frac{25}{2}z^2+\cdots .\notag 
\end{align} 
Using {\it Mathematica} you can get these $\Omega_i$ 
up to orders you need. It is also possible 
to derive these four periods directly  from Picard-Fuchs equation 
with aid of the software {\it Maple}. The command ``dsolve'' with 
options ``series'' and ``$z=0$'' gives you four independent 
series solutions up to orders you define, for example 
``Order $:=30$''. 

\subsection{Monodromy} 

In this model, there are seven singular points $z=\frac{1}{3},\
\frac{1}{4},\ \frac{1}{12},\ 0,\ -\frac{1}{6},\ -\frac{1}{5},\
-\frac{1}{4}$. It seems complicated to find complete monodromy behavior 
around every singular point. Since periods we have 
here are obtained by series expansion around the origin up to finite
orders, we can't anticipate the analytic property 
enough to determine the monodromy matrices, by 
continuation to other singular points. Then, following the literature \cite{ES}
, we have to determine the monodromy by numerical calculation 
with suitable approximations. 

The first step is to choose a reference point $p$ 
in items of each singular point. 
Next for each of the singular points $z_i$, 
we choose a piecewise linear loop starting and ending 
at the reference point $p$ 
and enclosing only one singular point.  
Using the {\it Maple} function ``dsolve'' with options ``numeric, 
method = gear, relerr= $10^{-15}$, abserr = $10^{-15}$'' and 
``Digits $:= 100$'', 
we can numerically integrate the differential
equation along these paths. 
Comparing integrated solutions to original ones at $p$
yields the monodromy matrices with respect to 
an arbitrary basis and produces fully filled
$4 \times 4$ matrices. The result will be recognized as 
the integer matrices with the precision of this calculation. 

\begin{figure}[h]
\begin{center}
\includegraphics[width=6cm,clip]{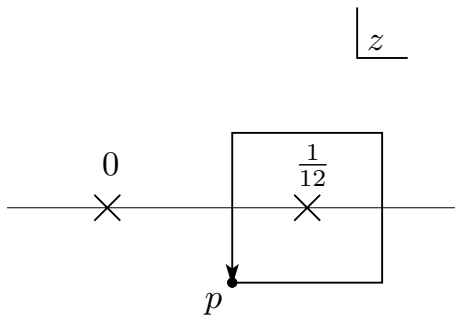}
\label{fig:one}
\caption{a loop from $p$ to $p$}
\end{center} 
\end{figure}

Symplectic basis we adopt here is 
\begin{align}
\omega^{P}_0&=\omega_0, \notag\\
\omega^{P}_1&=\frac{1}{2\pi i}\omega_1 \notag\\
\omega^{P}_2&=-\frac{c_2}{24}\,\omega_0+\alpha\,\frac{1}{2\pi i}\omega_1+
\frac{K}{2}\,\frac{1}{(2\pi i)^2}\omega_2, \\
\omega^{P}_3&=-\frac{\zeta(3)c_3}{(2\pi i)^3}\,\omega_0-\frac{c_2}{24}\,\frac{1}{2\pi i}\omega_1-
\frac{K}{6}\,\frac{1}{(2\pi i)^3}\omega_3, \notag
\end{align} 
where $K$ is Yukawa coupling, $c_2$ is second Chern class
\footnote{In mathematical literature, Yukawa coupling
$K$ is often denoted as $H^3$, and $c_2$ here is denoted as $c_2H$.},
$c_3=\chi$ is Euler number of Calabi-Yau manifold we consider here, 
and $\alpha$ is a constant 
which will be determined. So far this solution is  nothing but 
of the model on $X^{8,44}$ with Yukawa coupling 216, second Chern class
144, and Euler number $-72$. 
As is well known, there is an undetermined overall factor for $\omega_3^P$ here, and 
this would be fixed from the topological informations 
of Calabi-Yau manifold. In the models which admit freely 
acting group, these topological numbers will be reduced simultaneously
by moding out symmetries while their ratios will be kept. So we would like to use this degree of freedom
so that Euler number in this case will be reduced $\chi=-6$, 
and all monodromy matrices will be kept integral.  

With this basis, monodromy around the origin is given by 
\begin{align} 
M_{0}=\left(\begin{array}{crrrr}
1&0&0&0\\
1&1&0&0\\
\frac{K}{2}+\alpha&K&1&0\\
-\frac{c_2}{12}-\frac{K}{6}&-\frac{K}{2}+\alpha&-1&1\end{array} 
\right).
\end{align} 
A key combination of topological numbers to notice 
is $\frac{c_2}{12}+\frac{K}{6}$ when we reduce $\chi$. 
Also we choose $\alpha$ suitable so 
that elements of monodromy matrices will be integers, otherwise 
we set $\alpha=0$.

As the result, we can choose indices $K,\ c_2,\ c_3$ with
$\alpha=0$ as
\begin{gather}
K=18,\ c_2=12,\ c_3=-6, 
\end{gather} 
and monodromy matrices $M_{z_i}$ around $z=\frac{1}{3},\
\frac{1}{4},\ \frac{1}{12},\ 0,\ -\frac{1}{6},\
-\frac{1}{5},-\frac{1}{4}$ are, respectively, \newpage
\begin{align}&
\begin{pmatrix}
-11&-12&-12&48\\
3&4&3&-12\\
-3&-3&-2&-12\\
-3&-3&-3&13
\end{pmatrix},\ 
\begin{pmatrix}
-11&0&-12&72\\
2&1&2&-12\\
0&0&1&0\\
-2&0&-2&13
\end{pmatrix}, \
\begin{pmatrix}
1&0&0&12\\
0&1&0&0\\
0&0&1&0\\
0&0&0&1\end{pmatrix}, \notag
\\
&
\begin{pmatrix}
1&0&0&0\\
1&1&0&0&\\
9&18&1&0\\
-4&-9&-1&1
\end{pmatrix}, \ 
\begin{pmatrix}
-35&-96&-24&48\\
18&49&12&-24\\
-72&-192&-47&96\\
-27&-72&-18&37
\end{pmatrix},\ 
\begin{pmatrix}
-71&-180&-60&144\\
30&76&25&-60\\
-90&-225&-74&180\\
-36&-90&-30&73 \end{pmatrix}, 
\\
&\begin{pmatrix}
-35&-72&-36&108\\
12&25&12&-36\\
-24&-48&-23&72\\
-12&-24&-12&37
\end{pmatrix}. \notag
\end{align} 
Is is easy to check the consistency as 
\begin{gather}
M_{\frac{1}{3}}M_{\frac{1}{4}}M_{\frac{1}{12}}M_{0}M_{-\frac{1}{6}}
M_{-\frac{1}{5}}M_{-\frac{1}{4}}=\begin{pmatrix}
1&0&0&0\\
0&1&0&0\\
0&0&1&0\\
0&0&0&1\end{pmatrix}.
\end{gather} 
Above values of indices could be read from the analysis done by \cite{BCD}, however the conditions that matrix elements for monodromy have to be integral, 
appear to be able to determine these quantity. 

\subsection{Bi-linear form}

There is another way to find above indices by explicit evaluation of 
periods. Bi-linear form on periods $Bi(f,g)$ was invented 
as a tool to enumerate the symplectic relations among period integrals 
\cite{MS}. 


Let us consider some 
anti-symmetric differential operators $\partial^{k}\wedge \partial^{k'}$ 
acing to the solutions $f,\ g$  of 
Picard-Fuchs equation as 
\begin{gather}
\partial^{k}\wedge \partial^{k'} \,(f,g)=
{1 \over 2}(\partial^{k}f\cdot \partial^{k'}g-\partial^{k'}f\cdot
 \partial^kg)
\end{gather} 
where  $ \partial ^{k}$ are the $k$-th order differential operator 
with respect to moduli parameter. 
For periods $\{f_{\alpha_i},\ g_{\beta_j}\}$ obtained 
by integration along the symplectic homology basis
$\{\alpha_i,\ \beta_j\}$, we can make 
bi-liner form acting on these periods to have the 
same symplectic structure as homology cycles
\begin{gather}
Bi(f_{\alpha_i},g_{\beta_j})=-Bi(g_{\beta_j},f_{\alpha_i})=\delta_{i,j},\ 
Bi(f_{\alpha_i},f_{\alpha_j})=Bi(g_{\beta_i},g_{\beta_j})=0, 
\end{gather} 
up to normalization. This can be carried out by setting 
$Bi(f,g)$ to be some linear combination of $\partial^{k}\wedge
\partial^{k'}$, and imposing $\partial\, Bi(f,g)=0$ associated to 
Picard-Fuchs equation for periods.  
Using the ratios of coefficients of Picard-Fuchs equation,
we take $Bi(f,g)$ as 
\begin{gather}
Bi(f,g)=\exp\left(\frac{1}{2}\int r_3(z)dx\right)\left\{
\partial \wedge \partial^2\,(f,g)-1\wedge \partial^3\,(f,g)
-\frac{1}{2}r_3(z)\,1\wedge \partial^2\,(f,g) \right. \notag \\
\left.
+\left(\frac{1}{2}\partial r_3(z)+\frac{1}{4}r_3(z)^2-r_2(z)\right)
1\wedge \partial\,(f,g)
\right\}. 
\end{gather} 
Especially in the model, explicit evaluations around the origin 
show that 
\begin{gather}
Bi(\omega_0,\omega_3)=-2,\ Bi(\omega_1,\omega_2)=\frac{2}{3}, 
\end{gather} 
and all other combinations vanish. 

Using $Bi(f,g)$ on a solution around conifold point, 
we can estimate topological indices $c_1,\ c_2,\ K$ with Euler number 
$\tilde \chi$ of mirror manifold. 
In this model, conifold solutions around $z=\frac{1}{12}$ 
constitute of four 
kinds of functions whose leading behaviors 
$(z-\frac{1}{12})^{s}$ are of $s={0,1,1,2}$. We denote 
the polynomial solution with $s=1$ as $\omega_c$, which is
\begin{gather}
\omega_{c}=24u-288u^2+3264u^3-35712u^4+
\frac{1965312}{5}u^5-\frac{21731328}{5}u^6+\cdots 
\end{gather} 
where $u=z-\frac{1}{12}$. 
The topological indices can be obtained by using the ratio of 
bi-linear forms on a period $\omega_c$ and periods around the origin as
\begin{gather}
c_1=\frac{18\tilde\chi\zeta(3)}{\pi^2}\cdot 
\frac{Bi(\omega_1,\omega_c)}{Bi(\omega_3,\omega_c)},\notag\\
c_2=-\frac{18\tilde\chi\zeta(3)}{\pi^2}\cdot 
\frac{Bi(\omega_2,\omega_c)}{Bi(\omega_3,\omega_c)},\\ 
K=6\tilde\chi\zeta(3)\cdot 
\frac{Bi(\omega_0,\omega_c)}{Bi(\omega_3,\omega_c)}, \notag
\end{gather} 
with $\tilde \chi=-c_3$. Denominators are needed for correct 
normalization. Periods $w_i\ (i=0,1,2,3)$ behave well around the origin 
and bad around the conifold point $z=\frac{1}{12}$. Conversely, a period 
$\omega_c$ behaves bad around the origin and good around the conifold point.  
So we expect these quantities behave like constants in 
the intermediate region between the neighborhood of origin 
and the neighborhood of conifold point. 
With help of {\it Mathematica} or {\it Maple}, we can 
estimate the value of above expression by 
plotting from $z=0$ to $\frac{1}{12}$. 
Results for $c_2,\ K$ and $c_1$ with $\tilde \chi=-c_3=6$ are shown in fig.1, 
fig.2, and fig.3, respectively. 

\begin{figure}[h]
\begin{minipage}{5cm}
\begin{center}
\includegraphics[width=4.5cm,clip]{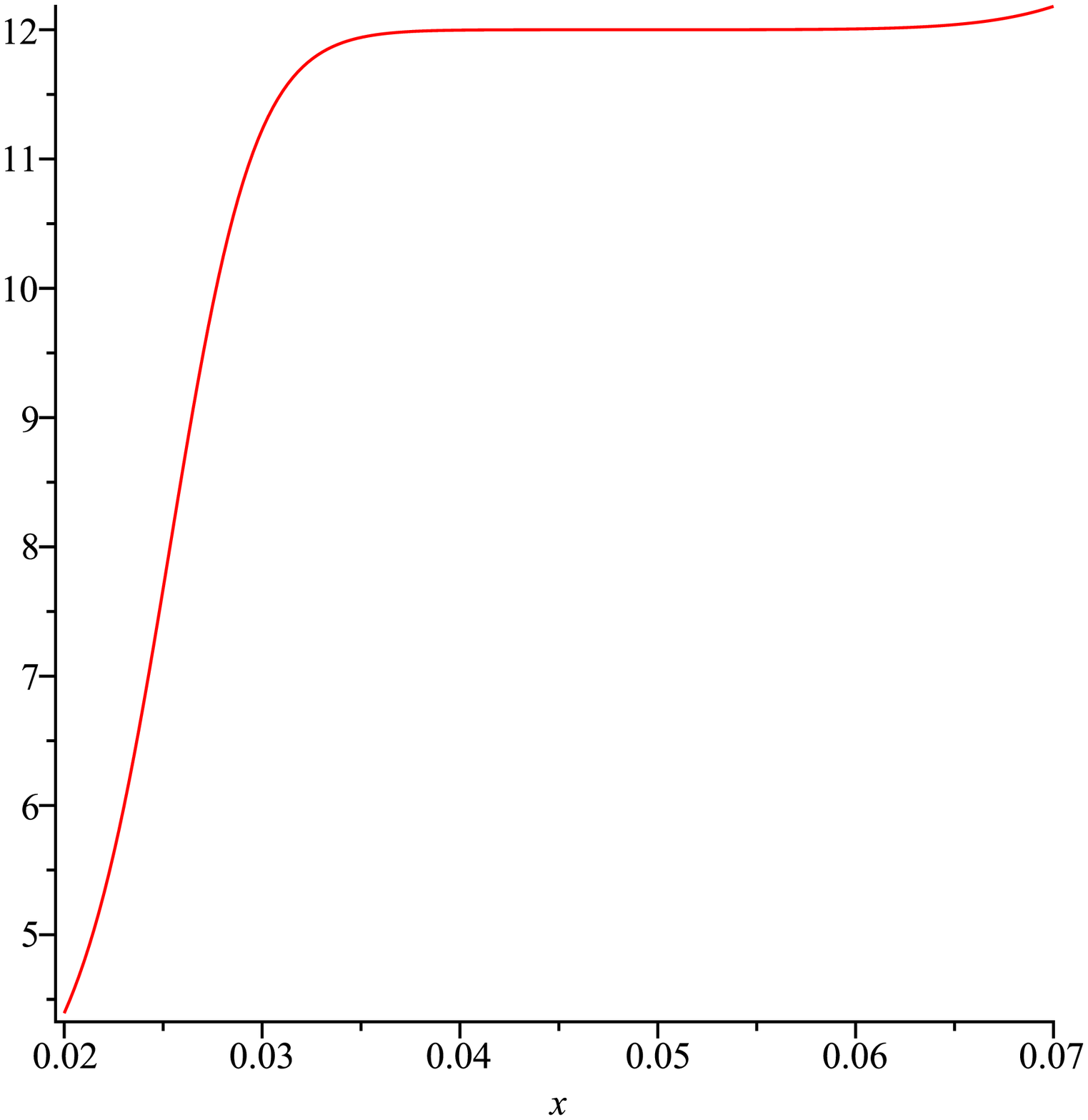}
\label{fig:two}
\caption{$c_2$}
\end{center} 
\end{minipage}
\begin{minipage}{5cm}
\begin{center}
\includegraphics[width=4.5cm,clip]{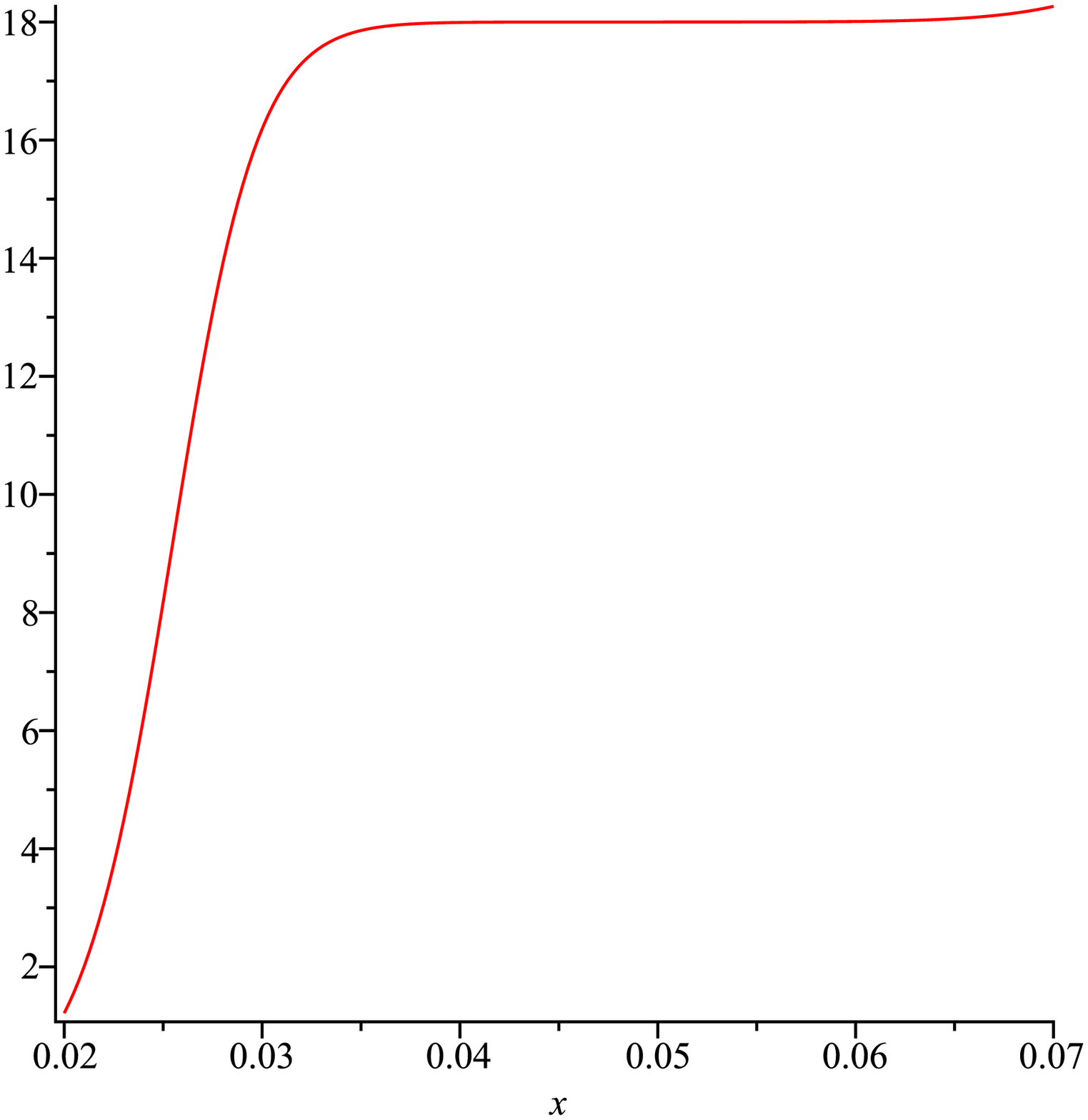}
\label{fig:three}
\caption{$K$}
\end{center} 
\end{minipage} 
\begin{minipage}{5cm}
\begin{center}
\includegraphics[width=4.5cm,clip]{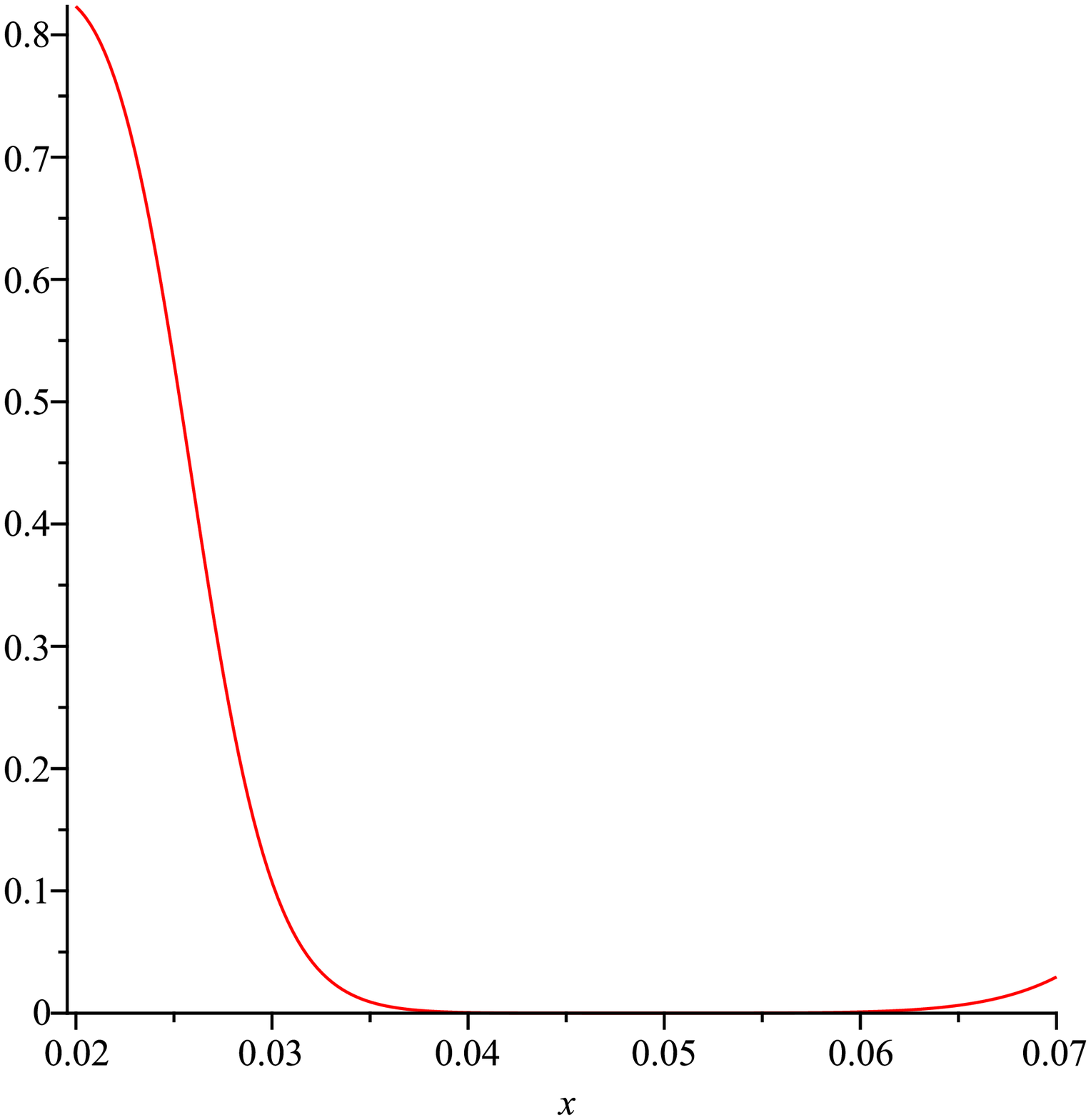}
\label{fig:four}
\caption{$c_1$}
\end{center} 
\end{minipage} 
\end{figure}
\noindent
The values of plateau parts of above results are  
same as the ones obtained from the monodromy matrices. 


\subsection{Instanton calculation}

Next we demonstrate the mirror symmetry to calculate 
the instanton numbers in genus 0 \cite{COGP} and 
genus 1 \cite{BCOV} in topological string theory. 
After compactification, we have 
following expansion for the partition function in  
topological string theory on Calabi-Yau manifold
\begin{gather}
F=\sum_{g=0}\lambda^{2g-2}F_{g}. 
\end{gather} 

For genus 0 case \cite{COGP}, we have the formula for the quantum Yukawa coupling 
as the triple derivative of free energy
\begin{gather}
\partial_t^3F_{0}=K+\sum_{i=1}\frac{a_i i^3q^i}{1-q^i}
\end{gather} 
where $a_i$'s are instanton numbers of the topological string
for genus 0. 
In order to calculate these instanton numbers we use the mirror
map. As usual, we set flat coordinate 
\begin{gather}
t=\dfrac{\omega_1(z)}{\omega_0(z)}=\log z+\frac{\Omega_1(z)}{\omega_0(z)}. 
\end{gather} 
We define  the variable $q=e^{t}$ and invert this relation to 
express $z$ in terms of $q$ as $z(q)$. Quantum Yukawa coupling 
is given by the the transformation of classical Yukawa coupling $K_c[z]$
from $z$ coordinate to $q$ as 
\begin{gather}
\partial_{t}^3F_0=
\dfrac{1}{z(q)^3}\left(\frac{\partial z(q)}{\partial q}\right)^3 
K_c[z(q)]\frac{1}{\omega_0(x(q))^2}
\end{gather} 
Equating (2.32) and (2.43), and expanding in terms of $q$, we find 
instanton number $a_i$. 

\begin{center}
$
\begin{array}{|r|r|}\hline 
i&a_i\\
\hline
1&6\\
2&15\\
3&30\\
4&114\\
5&522\\
6&2529\\
7&12636\\
8&69744\\
9&405168\\
10&2449773\\
11&15261150\\
12&97808574\\ \hline
\end{array} \ \ 
\begin{array}{|r|r|}\hline 
13&641284110\\
14&4287838548\\
15&29153498904\\
16&201163103922\\
17&1406107987374\\
18&9941935540692\\
19&71017384630734\\
20&511976000663130\\
21&3721663648494978\\
22&27257992426100979\\
23&201015705767041110\\
24&1491738880927589808\\
25&11134231701698352462\\
\hline
\end{array} 
$ \end{center} 

For genus 1 case, a derivative of free energy is the quantity we want to 
know
\begin{gather}
\partial_{t}F_1=-\frac{c_2}{24}+\sum_{i}(b_i+\frac{a_i}{12})\frac{q^i\cdot i}{1-q^i}
\end{gather} 
where $b_i$'s are instanton numbers of genus 1. To compute 
$\partial_tF_1$, we follow the analysis of holomorphic anomaly \cite{BCOV}, and 
use the formula
\begin{gather}
\partial_tF_1=-\frac{1}{2}\partial_t\log\left(
\frac{z^{1-\frac{c_2}{12}}Dis[z]}{\omega_0^{4-\frac{c_3}{12}}
\frac{\partial t}{\partial z}}
\right).
\end{gather} 
Here, we will use the an ansatz that $Dis[z]$ will be the weighted 
discriminant of the model. In well known examples with 
one moduli, we use a factor in the discriminant part as 
$(z-z_{c})^{-\frac{1}{6}}$ where $z_{c}$ are 
conifold point of the model, since the monodromy matrix around 
conifold point is usually set to be 
\[
 \begin{pmatrix}
1&0&0&1\\
0&1&0&0\\
0&0&1&0\\
0&0&0&1\end{pmatrix}.
\]
In this case we have different behavior around conifold as (2.19) because 
of effects by taking quotient of freely acting group. 
Our proposal for $Dis[z]$ in this case is the following 
\begin{gather}
Dis[z]=\prod_i(z-z_i)^{-\frac{\lambda_i}{6}}
\end{gather} 
where exponent $-\frac{\lambda_i}{6}$ will be determined for  
the monodromy matrix around the corresponding singular points 
by following method. 
Suppose $I$ is $4\times 4$ identity matrix, and $M_{z_i}$ 
is the monodromy matrix around $z_i$ where we take standard order of symplectic
basis 
as $(\omega_0^P,\ \omega_2^P,\ \omega_1^{P},\ \omega_3^P)^{T}$, then 
$M_{z_i}-I$ will be expressed 
by using a certain integral vector $(v_1,v_2,v_3,v_4)$ as 
\begin{gather}
M_{z_i}-I=-\lambda_i\,
\begin{pmatrix}
-v_4\\
v_3\\
-v_2\\
v_1
\end{pmatrix} 
\left(v_1,v_2,v_3,v_4\right)
\end{gather}
From this we can read off $\lambda_{i}$ for each $z_i$. 
In this model, we have
\begin{gather}
(z_i,\lambda_i)=(\frac{1}{12},12), 
(\frac{1}{4},2), (\frac{1}{3},3), (-\frac{1}{4},12), 
(-\frac{1}{5},1), (-\frac{1}{6},3), (0,0).
\end{gather} 
With these exponents, we can calculate instanton numbers in genus
1. Validity 
for the choice of $Dis[z]$ will be checked whether all $b_i$'s are 
integers or not. 

\begin{center}
$
\begin{array}{|rr|} \hline 
i&b_i\\ \hline 
1&7\\
2&41\\
3&233\\
4&1393\\
5&10121\\
6&72022\\
7&518960\\
8&3878 268\\
9&29 437 440\\
10&225 911 060\\
11&1 750 966 967\\
12&13 694 924 062\\
\hline
\end{array} \ \ 
\begin{array}{|r|r|}\hline 
13&107 873 536 349\\
14&854 827 410 657\\
15&6 809 292 590 762\\
16&54 489 457 053 320\\
17&437 778 784 226 585\\
18&3 529 641 546 245 282\\
19&28 547 903 108 757 361\\
20&231 550 298 613 514 152\\
21&1 882 881 825 812 617 783\\
22&15 346 314 478 913 958 426\\
23&125 342 659 401 860 309 785\\
24&1 025 721 457 879 954 913 034\\
25&8 408 667 562 177 554 413 449\\
\hline
\end{array} 
$
\end{center} 

Before closing this section, we mention that we can 
produce same results by using the original defining curve eq.(2.1) 
with reduced parameterization, say $c_1=c_2=c_3=0$, and with the 
period integral $\int\prod dt_ids_i\frac{1}{p\,q\,r}$.

\section{Mirror transformations of (1,5), (1,3), (1,1) model}

\subsection{Six manifolds with quaternionic symmetry}

In this section, we discuss three models with small Hodge numbers 
such as (5,1), (3,1), (1,1) by 
restricting the parameter of model coming from the manifolds $X^{4,68}$. 
For this manifold it is possible to write a defining 
polynomial that is transverse, as well as 
invariant and fixed point free under the group $\mathbb{H}\times 
\mathbb{Z}_2$, where $\mathbb{H}=
\left\{1,i,j,k,-1,-i,-j,-k\right\}$ is the quaternion group \cite{CD}.  
There are $3^4 = 81$ tetraquadric monomials in the 
$s_{\alpha}$ where $\alpha \in \mathbb{H}$. One of these is the fundamental
monomial, $\Pi_{\alpha \in \mathbb{H}}s_{\alpha}$, 
that is invariant under the full group. 
Of the other 80 monomials, 40 are even under $(s_\alpha,s_{-\alpha}) 
\rightarrow (s_\alpha,-s_{-\alpha})$ and 40 odd. 
The 40 even monomials fall into 
five parameter family of invariant polynomials. 
We change the variable as 
$s_1s_{-1}=t_1,\ s_is_{-i}=t_2,\ s_{j}s_{-j}=t_3,\ s_{k}s_{-k}=t_4$, 
five invariant polynomials are 

\begin{gather}
P_1=\left({\it t_1}\,{\it t_2}+{{1}\over{{\it t_1}\,{\it t_2}}}\right)
 \,\left({\it t_3}\,{\it t_4}+{{1}\over{{\it t_3}\,{\it t_4}}}\right)
 +\left({{{\it t_2}}\over{{\it t_1}}}+{{{\it t_1}}\over{{\it t_2}}}
 \right)\,\left({{{\it t_4}}\over{{\it t_3}}}+{{{\it t_3}}\over{
 {\it t_4}}}\right), \notag\\
P_2=\left({{{\it t_2}}\over{{\it t_1}}}+{{{\it t_1}}\over{{\it t_2}}}
 \right)\,\left({\it t_3}\,{\it t_4}+{{1}\over{{\it t_3}\,{\it t_4}}}
 \right)+\left({\it t_1}\,{\it t_2}+{{1}\over{{\it t_1}\,{\it t_2}}}
 \right)\,\left({{{\it t_4}}\over{{\it t_3}}}+{{{\it t_3}}\over{
 {\it t_4}}}\right), \notag\\
P_3=\left({\it t_3}+{{1}\over{{\it t_3}}}\right)\,\left({\it t_4}+{{1
 }\over{{\it t_4}}}\right)+\left({\it t_1}+{{1}\over{{\it t_1}}}
 \right)\,\left({\it t_2}+{{1}\over{{\it t_2}}}\right), \\
P_4=\left({\it t_2}+{{1}\over{{\it t_2}}}\right)\,\left({\it t_4}+{{1
 }\over{{\it t_4}}}\right)+\left({\it t_1}+{{1}\over{{\it t_1}}}
 \right)\,\left({\it t_3}+{{1}\over{{\it t_3}}}\right),\notag\\
P_5=\left({\it t_1}+{{1}\over{{\it t_1}}}\right)\,\left({\it t_4}+{{1
 }\over{{\it t_4}}}\right)+\left({\it t_2}+{{1}\over{{\it t_2}}}
 \right)\,\left({\it t_3}+{{1}\over{{\it t_3}}}\right).\notag
\end{gather} 
Manifolds which are defined by these polynomials have been found 
in \cite{CD}. 
$X^{4,68}$ modulo $\mathbb{H}\times \mathbb{Z}_2$ is a model 
with Hodge numbers (1,5), because $\chi=-128/(8\times 2)=-8$, 
which is smallest combination of Hodge 
numbers constructed on this manifold. 
The model we consider here are given by defining polynomial as 
\begin{gather}
 f=1+\sum_{i=1}^5 c_iP_i. 
\end{gather} 

In this section, we pursuit the possibility that the
mirror model of (1,$h^{2,1}$) with $h^{2,1}\leq 5$ 
would be obtained by suitable restriction of parameter $c_i$'s of 
the above curve reducing the number of moduli to 1. It is natural 
to get the mirror of (1,5) model with above curve because it is 
just the restriction of defining curve of (1,5). 
Besides this case, it is interesting to get mirror models of (1,3) 
and (1,1) with the family of this curves 
associated with invariant polynomial on $X^{4,68}$.

\subsection{Mirror transformation of (1,5)} 

First we consider the model defined by one of $P_i$'s, for example, 
\begin{gather}
f=1+cP_3. 
\end{gather} 
We denote this model as (5,1) because the indices we find  
below will be $\tilde \chi =8$. Fundamental period $\omega_0$ of
this model is calculated in the same way as in previous section.  
It is easy see that the curve with any $P_i$ will produce the 
same fundamental period. 
We have 
Picard-Fuchs equation for periods, whose coefficients are given by
\begin{gather}
a_4(z)=z^3(64z-1)(16z-1), \notag\\
a_3(z)=6z^2-640z^3+10240z^4, \notag \\
a_2(z)=7z-1172z^2+25344z^3,\\
a_1(z)=1-424z+14592z^2,\notag\\
a_0(z)=-8+768 z.\notag
\end{gather}
This Picard-Fuchs equation is ``AESZ 16'' in the database \cite{AESZ}. 
The local property of the solution is given by  
\begin{align}
\left\{
\begin{array}{cccc}
0&1/64&1/16&\infty\\
\hline
0&0&0&\frac{1}{2}\\
0&1&1&1\\
0&1&1&1\\
0&2&2&\frac{3}{2}
\end{array}
\right\}
\end{align} 
Monodromy matrices are found as following form with $\alpha=\frac{1}{2}$
\begin{align}
&M_{\frac{1}{64}}=\left(\begin{array}{rrrr}
1&0&0&16\\
0&1&0&0\\
0&0&1&0\\
0&0&0&1\end{array} 
\right),\ \ \ 
M_{0}=\left(\begin{array}{rrrr}
1&0&0&0\\
1&1&0&0\\
2&3&1&0\\
-1&-1&-1&1\end{array} 
\right),\notag\\
&M_{\frac{1}{16}}=\left(\begin{array}{rrrr}
-7&8&-16&64\\
2&-1&4&-16\\
1&-1&3&-8\\
-1&1&-2&9\end{array} 
\right). 
\end{align} 
Monodromy around $\infty$ is not trivial in this case, so including this 
contribution we can check the consistency as
$M_0M_{1/64}M_{1/16}M_{\infty}=I$. 
With this result we see that indices of this model are 
\begin{gather}
K=3,\ c_2=6,\ -\tilde\chi=c_3=-8. 
\end{gather} 
These are checked directly by using bi-linear form $Bi(f,g)$.  

Using classical Yukawa coupling 
\begin{gather}
K_c[z]=\frac{3}{(1-16z)(1-64z)}
\end{gather} 
we can calculate genus 0 instanton numbers $a_i$. 

\begin{center}
$\begin{array}{|r|r|}\hline 
i&a_i\\ \hline 
1&12\\
2&60\\
3&644\\
4&9216\\
5&157536\\
6&3083604\\
7&66250884\\
8&1522656816\\
9&36850292240\\
10&929119768416\\
11&24217533456516\\
12&648807231571968\\
\hline
\end{array} \ \ 
\begin{array}{|r|r|} \hline 
13&17788009827334944\\
14&497375443061477076\\
15&14145255850235272728\\
16&408279490665349434096\\
17&11938435093860094144356\\
18&353131094729321849805456\\
19&10553174109736271978455644\\
20&318296315795274110349024768\\
21&9 680 349 870 962 148 118 941 442 064\\
22&296 637 049 016 097 525 560 121 350 484\\
23&9 152 575 814 156 431 319 768 582 760 780\\
24&284 178 186 604 373 405 325 304 089 538 064\\
25&8 874 513 455 364 110 811 718 556 119 122 168\\
\hline
\end{array} 
$
\end{center}

From the monodromy matrices, we also have the exponents of 
singular points of the weighted discriminant for the genus 1 free energy
\begin{gather}
\log dis[z]=-\frac{1}{6}\left\{\log(1-16z)+16\log(1-64z)\right\}. 
\end{gather} 

Genus 1 instanton numbers $b_i$ are

\begin{center}
$\begin{array}{|r|r|} \hline 
i&b_1\\ \hline 
1&62\\
2&  944\\
3&  23 418\\
4&  617 181\\
5&  17 548 056\\
6&  519 858 096\\
7&  15 862 425 890\\
8&  494 434 015 977\\
9&  15 664 741 989 264\\
10&  502 692 670 367 672\\
11&  16 299 569 600 798 954\\
12&  533 034 159 955 717 583\\
\hline
\end{array} \ \ 
\begin{array}{|r|r|} \hline 
13&  17 556 644 191 335 142 120\\
14&  581 797 187 945 487 716 872\\
15&  19 381 087 602 033 473 111 548\\
16&  648 582 779 031 746 543 753 753\\
17&  21 791 736 751 065 456 207 143 258\\
18&  734 780 312 858 320 227 411 506 408\\
19&  24 853 931 064 859 179 585 102 463 230\\
20&  843 069 010 900 057 413 932 047 054 080\\
21&  28 670 876 616 718 121 788 361 190 305 616\\
22&  977 293 230 328 375 433 391 862 511 358 528\\
23&  33 382 975 124 863 651 007 753 794 577 850 158\\
24&  1 142 518 107 949 648 008 219 538 769 541 941 587\\
25&  39 171 513 766 427 556 729 463 448 532 706 463 028\\
\hline
\end{array} 
$
\end{center}

\subsection{A candidacy model as a mirror of (1,3)} 

Next we consider the mirror model whose defining curve is made out of 
two kinds of invariant polynomials $\{P_i\}$. 
We anticipate to have some extra symmetry $\mathbb{Z}_2$ to 
reduce Euler number from $-8$ to $-4$ by combining $\{P_i\}$. 
Recently in \cite{CCM} the (1,3) model has been discovered 
as a quotient $\mathbb{H}\times \mathbb{Z}_2\times \mathbb{Z}_2$ 
on $X^{1,65}$, which may be related to the model we are going to
construct here. 
The criteria to adopt a combination for defining curve as 
the mirror model of (1,3) are following; 

1. derived Picard-Fuchs equation which is satisfied by calculated 
period integral is of 4th order equation and of Calabi-Yau type. 

2. assuming Euler number $\chi=-4$, monodromy matrices are all integral
and consistent. 

\noindent
For defining the model there would be several possibilities. For
example,  
combinations such as $\{P_1,P_2\},\ \{P_3,P_4\},\ \{P_3,P_5\},\ \{P_4,P_5\}$ 
would become Calabi-Yau manifolds whose Hodge number could not 
be small. 

A choice we take  here is 
\begin{gather}
 f=1+c(P_1+2P_3+8).   
\end{gather} 
Due to the relation
\begin{gather}
P_1+2P_3+8=
\left({\it t_1}\,{\it t_2}+{{1}\over{{\it t_1}\,{\it t_2}}}+2\right)
 \,\left({\it t_3}\,{\it t_4}+{{1}\over{{\it t_3}\,{\it t_4}}}+2\right)
 +\left({{{\it t_2}}\over{{\it t_1}}}+{{{\it t_1}}\over{{\it t_2}}}
 +2\right)\,\left({{{\it t_4}}\over{{\it t_3}}}+{{{\it t_3}}\over{
 {\it t_4}}}+2\right), 
\end{gather} 
and by changes of variables, this curve is expressed as follows
\begin{gather}
f=1+c\left((\zeta_1+\frac{1}{\zeta_1})^2(\zeta_2+\frac{1}{\zeta_2})^2
+(\xi_1+\frac{1}{\xi_1})^2(\xi_2+\frac{1}{\xi_2})^2
\right).
\end{gather} 
The quadratic form of this curve may enhance the symmetry of the models. 
We do not investigate here singularities and 
degrees of freedom of deformations corresponding to this model in detail, 
we anticipate that three polynomials 
$\{P_1+2P_i+8\}$ $(i=3,4,5)$ would define $(1,3)$ Calabi-Yau space somehow. 
We would like to refer the model defined by eq.(3.47) 
as the mirror of (1,3) due to Euler number 
$\tilde \chi=-c_3$ coming from the monodromy as we will see below. 

Picard-Fuchs equation is expressed by following coefficients
\begin{align}
a_4(z)=&z^3(1-32 z)(16 z-1)^2(32z-3)^2, \notag\\
a_3(z)=&-2 z^2(16 z-1)(32 z-3)(81920 z^3-14336 z^2+688 z-9),\notag\\
a_2(z)=&-z(-63+9132 z
-410240 z^2+7860224 z^3-66977792 z^4+209715200 z^5),\\
a_1(z)=&9-3000 z+188928 z^2
-4259840 z^3+38797312 z^4-125829120 z^5,\notag\\
a_0(z)=&-(512 z-24)(16384 z^3-4096 z^2+336 z-3).\notag
\end{align}
This operator is found as ``AESZ 23'' in the database \cite{AESZ}. 
The local properties of the solutions are read as 
\begin{align}
\left\{
\begin{array}{ccccc}
0&1/32&1/16&3/32&\infty\\
\hline
0&0&0&0&1\\
0&1&1/2&1&1\\
0&1&1/2&3&1\\
0&2&1&4&2
\end{array}
\right\}
\end{align} 
Using bi-linear form, we 
first estimate the relation between $K,\ c_2$ and $c_3$ numerically
\begin{gather}
K=-3c_3,\ \ c_2=-3c_3. 
\end{gather}
Next we analyze monodromy behavior around $z=0,\ \frac{1}{32},\
\frac{1}{16}$ with unknown $c_3$ as
\begin{align}
&M_{\frac{1}{32}}=\left(\begin{array}{rrrr}
1&0&0&-\frac{32}{c_3}\\
0&1&0&0\\
0&0&1&0\\
0&0&0&1\end{array} 
\right),\ \ 
M_{0}=\left(\begin{array}{rrrr}
1&0&0&0\\
1&1&0&0\\
-\frac{3}{2}c_3&-3c_3&1&0\\
-\frac{3}{4}c_3&\frac{3}{2}c_3&-1&1\end{array} 
\right),\notag\\
&M_{\frac{1}{16}}=\left(\begin{array}{rrrr}
1&-8&0&\frac{64}{c_3}\\
-1&3&\frac{4}{c_3}&-\frac{32}{c_3}\\
-\frac{c_3}{2}&0&3&-8\\
-\frac{c_3}{4}&\frac{c_3}{2}&1&-7\end{array} 
\right), 
\end{align} 
where we set $\alpha=0$. 
This result shows that indices for this models must be 
\begin{gather}
c_3=-4,\  K=12,\ c_2=12, \ \tilde \chi=4.  
\end{gather} 
This is the reason why we refer this model as the mirror of (1,3). 

%

One of strange things in this model is that monodromy matrix 
around $z=\frac{1}{16}$ does not have standard form, and 
its square becomes the one we expected 
\begin{gather}
M_{\frac{1}{16}}^2=\left(\begin{array}{rrrr}
-7&0&-8&32\\
2&1&2&-8\\
0&0&1&0\\
-2&0&-2&9\end{array} 
\right).
\end{gather}

Classical Yukawa coupling is 
\begin{gather}
K_c[z]=\frac{4(3-32z)}{(1-16z)^2(1-32z)},
\end{gather} 
therefore we have genus 0 instanton numbers $a_i$ as 

\begin{center}
$\begin{array}{|r|r|} \hline 
i&a_i\\ \hline 
1& 16\\
2& 52\\
3& 176\\
4& 1112\\ 
5& 9344\\
6& 79 420\\
7& 735 408\\ 
8& 7 426 680\\ 
9& 78 932 672\\ 
10& 871 171 744\\
11& 9 941 092 528\\ 
12& 116 637 185 736\\ 
\hline
\end{array} \ \ 
\begin{array}{|r|r|} \hline 
13& 1 400 292 243 840\\
14& 17 143 914 174 524\\ 
15& 213 484 884 447 264\\ 
16& 2 697 997 276 049 144\\
17& 34 542 569 526 333 232\\ 
18& 447 367 338 559 094 512\\
19& 5 853 717 440 568 978 896\\ 
20& 77 303 742 587 189 634 752\\
21& 1 029 391 625 209 701 923 520\\ 
22& 13 811 360 059 974 386 521 148\\
23& 186 585 097 601 501 528 151 824\\ 
24& 2 536 577 821 151 895 406 785 576\\
25& 34 683 844 685 464 450 280 801 952\\ 
\hline
\end{array} 
$
\end{center} 

For genus 1 case, we have to get the exponents of singular points of 
$Dis[z]$ from monodromy matrices. As we have mentioned, monodromy around 
$z=\frac{1}{16}$ are not usual, so we propose that logarithm of $Dis[z]$ 
would be 
\begin{gather}
\log Dis[z]=-\frac{1}{6}\left\{\lambda_1\log(1-16z)^2+
\lambda_2\log(1-32z)\right\}, 
\end{gather} 
where $\lambda_1$ is the exponent from $M_{\frac{1}{16}}^2$. 
With exponents $\lambda_1=2,\ \lambda_2=8$, 
the result for genus 1 instanton calculation is 



\begin{center}
$\begin{array}{|r|r|} \hline 
i&b_i\\ \hline 
1& 8\\ 
2& 82\\ 
3& 856\\ 
4& 10 321\\ 
5& 128 864\\
6& 1 677 110\\ 
7& 22 506 040\\ 
8& 308 025 697\\ 
9& 4 282 495 040\\
10& 60 292 530 504\\ 
11& 857 470 990 104\\ 
12& 12 296 761 625 755\\
\hline
\end{array} \ \ 
\begin{array}{|r|r|} \hline 
13& 177 583 895 318 624\\ 
14& 2 579 924 022 491 086\\
15& 37 674 030 557 685 648\\ 
16& 552 612 289 406 933 025\\
17& 8 137 788 233 521 859 928\\ 
18& 120 255 275 347 028 012 752\\
19& 1 782 584 199 002 075 687 048\\ 
20& 26 497 544 221 536 391 150 192\\
21& 394 868 645 365 328 468 752 512\\ 
22& 5 897 791 770 766 814 586 563 334\\
23& 88 272 996 476 471 800 976 727 752\\ 
24& 1 323 704 678 176 645 827 390 945 547\\
25& 19 884 364 383 541 833 676 997 550 064\\

\hline
\end{array} 
$
\end{center} 

Before closing this subsection we mention about results given by 
similar calculations based on mirror transformations 
in two different models known as (1,3) 
 in \cite{CD, BDS}. First model is a quotient 
$\mathbb{Z}_{10}\times \mathbb{Z}_2$ constructed on $X^{5,45}$. 
Following the literature,
defining curve for a mirror of this model would be 
\begin{align}
p_1=&1+a_1(t_4t_5+t_5t_1+t_1t_2+t_2t_3+t_3t_4)
+a_2(t_3t_5+t_4t_1+t_5t_2+t_1t_3+t_2t_4) \notag\\
&+a_3t_1t_2t_3t_4t_5(\frac{1}{t_1}+
\frac{1}{t_2}+\frac{1}{t_3}+\frac{1}{t_4}+\frac{1}{t_5}), \notag\\
p_2=&t_1t_2t_3t_4t_5+a_1(t_1t_2t_3+t_2t_3t_4+t_3t_4t_5+
t_4t_5t_1+t_5t_1t_2) \notag \\
&+a_2(t_1t_2t_4+t_2t_3t_5+t_3t_4t_1+
t_4t_5t_2+t_5t_1t_3)+a_3(t_1+t_2+t_3+t_4+t_5). 
\end{align} 
with restriction $a_1=a_2=0$. We can derive Picard-Fuchs equation 
(AESZ 34 \cite{AESZ}), and expect consistent result for 
$\chi=-4$ with $K=6,\ c_2=6$. However, 
monodromy matrices can not be integral, thus we would conclude 
this is a mirror of (1,5) model of $\chi=-8$ with $K=12,\ c_2=12$. 

Second one is the model on $X^{19,19}$ modulo 
$\mathrm{Disc}_3\cong\mathbb{Z}_3\rtimes\mathbb{Z}_4$. In the literature
, explicite curve to define a mirror of this model is not found, so 
we propose following form 
\begin{align}
p&=1+s_0s_1+s_1s_2+s_2s_0,\ \ q=1+t_0t_1+t_1t_2+t_2t_0,\notag\\
r&=s_0s_1s_2t_0t_1t_2+c(s_0t_0+s_1t_1+s_2t_2+s_0t_1+s_1t_2+
s_2t_0+s_0t_2+s_1t_0+s_2t_1). 
\end{align}
Calculations lead us to Picard-Fuchs equation (AESZ 103 \cite{AESZ}), 
and results about monodromy and instanton calculations 
done by mirror transformation are all 
consistent for $\chi=-4$ with $K=3,\ c_2=6$. 

\subsection{Mirror transformation of minimal model (1,1)} 

The model which has minimal Hodge numbers is (1,1). 
This model is originally found by studying 24-cell in \cite{Br}. 
A example of curve to define this model in $\mathbb{C}^8$ 
is 
\begin{gather}
p=1+\sum_{i=1}^8x_i+x_1x_3+x_1x_5+x_2x_6+x_3x_7+x_3x_5+x_6x_8+x_1x_3x_7 
\notag \\
+x_3x_6(x_1+x_3+x_4+x_5+x_6+x_7+x_8+x_3x_6)+\varphi x_3x_6,
\end{gather} 
with identifications 
\begin{gather}
x_1x_3=x_2x_4,x_1x_5=x_4x_6,x_1x_7=x_2x_6,x_1x_8=x_2x_5=x_3x_6=x_4x_7,
\notag \\
x_2x_8=x_3x_7,x_3x_5=x_4x_8,x_5x_7=x_6x_8.
\end{gather} 
Reducing the number of variables 
by using above identifications from eight to four, and 
changing variables, effective curve would be 
\begin{align}
f=1+&c\left(s_1+s_2+s_3+s_4+\frac{1}{s_1}+\frac{1}{s_2}+\frac{1}{s_3}
+\frac{1}{s_4} \right.\notag \\
&+s_2s_3+\frac{1}{s_2s_3}+\frac{s_1}{s_4}+\frac{s_4}{s_1}+\frac{s_1}{s_2}+
\frac{s_2}{s_1}+\frac{s_3}{s_4}+\frac{s_4}{s_3} \\
&\left. + \frac{s_1}{s_2s_3}+\frac{s_2s_3}{s_1}+
\frac{s_1}{s_2s_4}+\frac{s_2s_4}{s_1}+\frac{s_4}{s_2s_3}+
\frac{s_2s_3}{s_4}+\frac{s_4}{s_1s_3}+\frac{s_1s_3}{s_4}
\right), \notag
\end{align} 
where moduli $c=1/\varphi$

As this minimal model (1,1), 
we propose alternative  
definition of curve made out of invariant polynomials $\{P_i\}$ on $X^{4,48}$
\begin{gather}
 f=1+c(P_3+P_4+P_5). 
\end{gather}
This is unique definition of using three kinds of invariant 
polynomials with $Z_3$ symmetry. 
This definition is different from the curve (3.78),  
however we can show that periods, 
monodromy matrices, and 
instanton numbers obtained by both definitions are completely same. 

From the series expansion of fundamental period, we have 
Picard-Fuchs equation of the form
\begin{align}
a_4(z)=&-z^3(8 z+1)(24 z-1)(3 z+1)(4 z+1)(12 z+1)(1+18 z)^2,\notag\\
a_3(z)=& 6 z^2+204 z^3-1948 z^4-184248 z^5-3322944 z^6-26476416 z^7
\notag\\
& -95551488 z^8-125411328 z^9,\notag\\
a_2(z)=&7 z+164 z^2-8310 z^3-455148 z^4-8595936 z^5
-77054976 z^6\notag \\
& -319997952 z^7-483729408 z^8,\\
a_1(z)=&1-14 z-5574 z^2-274788 z^3-5818176 z^4-60943104 z^5\notag \\
&-300589056 z^6-537477120 z^7,\notag\\
a_0(z)=&-384 z-22752 z^2-606528 z^3-7921152 z^4-48771072 z^5\notag \\
&-107495424 z^6.\notag
\end{align}
This is nothing but ``AESZ 366'' in the database \cite{AESZ}. 
The local property for periods is summarized as 
\begin{align}
\left\{
\begin{array}{cccccccc}
-1/3&-1/4&-1/8&-1/12&-1/18&0&1/24&\infty\\
\hline
0&0&0&0&0&0&0&1\\
1&1&1&1&1&0&1&2\\
1&1&1&1&3&0&1&2\\
2&2&2&2&4&0&2&3
\end{array}
\right\}
\end{align} 
Monodromy matrices are found as 
\begin{align}
&M_{\frac{1}{24}}=\left(\begin{array}{rrrr}
1&0&0&24\\
0&1&0&0\\
0&0&1&0\\
0&0&0&1\end{array} 
\right),
M_{0}=\left(\begin{array}{rrrr}
1&0&0&0\\
1&1&0&0\\
2&4&1&0\\
-1&-2&-1&1\end{array} 
\right),\notag\\
&M_{-\frac{1}{12}}=\left(\begin{array}{rrrr}
1&0&0&0\\
1&3&1&0\\
-2&-4&-1&0\\
-1&-2&-1&1\end{array} 
\right),
M_{-\frac{1}{8}}=\left(\begin{array}{rrrr}
-23&-48&-48&72\\
16&33&32&-48\\
-16&-32&-31&48\\
-8&-16&-16&25\end{array} 
\right),\\
&M_{-\frac{1}{4}}=\left(\begin{array}{rrrr}
-95&-144&-264&576\\
44&67&121&-264\\
-24&-36&-65&144\\
-16&-24&-44&97\end{array} 
\right),
M_{-\frac{1}{3}}=\left(\begin{array}{rrrr}
-95&-96&-288&768\\
36&37&108&-288\\
-12&-12&-35&96\\
-12&-12&-36&97\end{array} 
\right). \notag
\end{align} 
From these matrices, we would read indices of this models with
$\alpha=0$ as 
\begin{gather}
K=4,\ c_2=4,\ c_3=-\tilde \chi=0.
\end{gather} 
Apart from previous examples, we are not able to check by bi-linear 
form $Bi(f,g)$ in this case, because $\tilde \chi=0$. 

Curious results appear about genus 0 instanton numbers. Following same 
procedure as before, and using classical Yukawa coupling
\begin{gather}
K_c[z]=\frac{4(1+18z)}{(1+3z)(1+4z)(1+8z)(1+12z)(1-24z)}, 
\end{gather} 
we find that instanton numbers in genus 0 level 
for even order become negative. 

\begin{center}
$\begin{array}{|r|r|} \hline 
i&a_i\\ \hline 
1& 12\\ 
2& -16\\ 
3& 256\\ 
4& -1012\\ 
5& 17 168\\ 
6& -102432\\
7& 1 768032\\ 
8& -12 810048\\ 
9& 226260008\\ 
10& -1 831410544\\
11& 33 000429000\\ 
12& -286340050052\\ 
\hline
\end{array} \ \ 
\begin{array}{|r|r|} \hline
13& 5 252822116016\\
14& -47 718467477584\\ 
15& 890108488876160\\ 
16& -8 340130846927456\\
17& 158096635640838140\\ 
18& -1 512328959263997360\\
19& 29 129403982340313132\\ 
20& -282368793768124234092\\
21& 5 527396080871599103212\\ 
22& -53 986928091516821971440\\
23& 1 074486987800843943995916\\ 
24& -10 525957761076292523611520\\
25& 213137290904593560452816768\\ 
\hline
\end{array} 
$
\end{center}

For genus 1 instanton numbers, we have to find 
correct exponents of $Dis[z]$. Direct calculations about monodromy 
matrices lead us the form 
of $\log Dis[z]$ as
\begin{align}
\log Dis[z]=-\frac{1}{6}\{
12\log(1+3z)+&\log(1+4z)+8\log(1+8z)\\
&+24\log(1-24z)+\log(1+12z)\}\notag
\end{align}
Differently from the previous
models, this discriminant produces wrong genus 1 behaviors whose
instanton numbers are half-integers, such as 
\begin{gather}
b_1=35,\ b_2=\frac{753}{2},\ b_3=3175,\ b_4=45510,\ b_5=501917,\
b_6=\frac{1583609}{2}, \ \cdots
\end{gather} 
Surprisingly, if we set the coefficient of $\log(1+12z)$ in $\log Dis[z]$ to
be 5 mod 6, every genus 1 instanton number becomes integer
 up to 50th orders. 

Lastly we add some comments about results 
of a model made of another combination of three 
$P_i$, whose defining curve is $f=1-c(P_1+P_2+4P_3)$, though this 
is not relevant to the model (1,1). Picard-Fuchs
equation of this model looks ordinary Calabi-Yau type (
AESZ 107 \cite{AESZ}), however 
topological indices turn out to be unusual values 
as $K=4,\ c_2=4,\ \chi=10$. Monodromy matrices are consistent, and 
instanton numbers are all integers in genus 0 and 1 level, 
however some of them become negative.

\section{Conclusion and Discussions}\label{conclusion}
We have presented mirror transformations of 
Calabi-Yau manifolds whose Hodge numbers $(h^{11},h^{21})$ are both
small. We have determined the monodromy of the models completely, 
and enumerated genus 0 and 1 instanton 
numbers of the models by using weighted discriminant 
for genus 1 level. 
Results based on the mirror models of (1,5) and (1,4) are 
consistent. Since the Yukawa coupling as well as instanton numbers in genus 0 in
these quotient models are directly related to the quantities  
on originated manifolds by division of freely acting group,  
genus 1 calculations are more significant. 
We have also proposed the description 
for mirrors of (1,3) and (1,1) models by using 
invariant polynomials of (1,5) model. 
Results in (1,3) case look reasonable, however 
in minimal case (1,1) negative and half integer value 
of instanton numbers appear against our expectation. 
Special treatment might be needed when 
you calculate instanton numbers in the model
with Euler number $\chi \geq 0$. 

We attempted as many combinations as possible of invariant polynomials 
that could be viewed as definitions of mirror models 
of Calabi-Yau with small Hodge numbers discussed in section 3, 
however we couldn't find an appropriate one to a mirror of (1,2) model.  
The extention to include sets of invariant polynomials of 
(1,4) model on $X^{8,44}$, or (1,3) model on $X^{5,45}$ did not
work well so far. It is interesting to recognize how to describe the model (1,2) 
in a way suggested in \cite{SW} 
and its mirror.  

The numerical integration around singular points to fix 
monodromy behaviors would be applicable to several modulus case. 
This method may help us to perform mirror transformations 
for various string compactifications. 
It is interesting 
if $(2,2)$ model could be analyzed in a view point of mirror symmetry 
transformation by applying methods we discussed here, 
as well as the conifold transition to other models such as 
(1,3) and (1,4). 

Also investigations to apply these methods to the mirror symmetry 
with small Hodge numbers in open string theories including D-branes
would be interesting.



\end{document}